\documentstyle[amssymb,twocolumn]{article}

\makeatletter
\def\@normalsize{\@setsize\normalsize{12pt}\xpt\@xpt
\abovedisplayskip 10pt plus2pt minus5pt\belowdisplayskip \abovedisplayskip
\abovedisplayshortskip \z@ plus3pt\belowdisplayshortskip 6pt plus3pt
minus3pt\let\@listi\@listI}

\def\subsize{\@setsize\subsize{12pt}\xipt\@xipt}

\def\section{\@startsection {section}{1}{\z@}{12pt plus 2pt minus 2pt}
{12pt plus 2pt minus 2pt}{\large\bf}}

\def\subsection{\@startsection {subsection}{2}{\z@}{12pt plus 2pt minus 2pt}
{12pt plus 2pt minus 2pt}{\subsize\bf}}

\def\@maketitle{\newpage
 \null
 \begin{center}%
  \vspace*{.125in}
  {\Large \bf \@title \par}%
  \vspace*{\baselineskip}
  \vspace*{\baselineskip}
  {\large
   \lineskip .6em
   \begin{tabular}[t]{c}\@author
   \end{tabular}\par}%
  \vspace*{\baselineskip}
  {\large \@date}%
 \end{center}%
 \par
 \vspace*{\baselineskip}
 }

\@addtoreset{equation}{section}

\def\@biblabel#1{#1.}

\def\thebibliography#1{\section*{\refname\@mkboth
  {\uppercase{\refname}}{\uppercase{\refname}}}\list
  {\@biblabel{\arabic{enumiv}}}{\settowidth\labelwidth{\@biblabel{#1}}%
    \leftmargin\labelwidth
    \advance\leftmargin\labelsep
    \usecounter{enumiv}%
    \let\p@enumiv\@empty
    \def\theenumiv{\arabic{enumiv}}}%
    \def\newblock{\hskip .11em plus.33em minus.07em}%
    \sloppy\clubpenalty4000\widowpenalty4000
    \sfcode`\.=1000\relax}

\makeatother

\newcommand{\ket}[1]{{|#1\rangle}}

\newcommand{\calC}{{\cal C}}

\begin{document}
\bibliographystyle{alpha}

\renewcommand{\thefootnote}{*}

\title{ Fault-Tolerant Quantum Computation}
\author{Peter W. Shor\\
AT\&T Research\\
Room 2D-149\\
600 Mountain Ave.\\
Murray Hill, NJ \  07974, USA \\
shor@research.att.com
}
\date{}
\maketitle
\pagestyle{plain}
\section*{\centering Abstract}
{
\it
It has recently been realized that use of the properties of quantum 
mechanics might speed up certain computations dramatically.
Interest in quantum computation has since been growing.
One of the main difficulties in realizing quantum computation is
that decoherence tends to destroy the information in a superposition of
states in a quantum computer, making long computations
impossible.  A further difficulty is that inaccuracies in quantum state 
transformations throughout the computation accumulate, rendering 
long computations unreliable.  However, these obstacles may not be
as formidable as originally believed.  For any
quantum computation with $t$ gates, we show how to build a polynomial size
quantum circuit that tolerates O($\mathit 1/\log^c t$) 
amounts of inaccuracy and decoherence per gate, for some constant $c$;
the previous bound was O($\mathit 1/t$).
We do this by showing that operations can be performed on quantum data
encoded by quantum error-correcting codes without decoding this data.
}

\vspace{\baselineskip}

\sloppy

\setcounter{page}{1}

\section{Introduction}

It has recently been discovered that certain properties of quantum
mechanics have a profound effect on abstract models of
computation.  More specifically, by using the superposition and 
the interference principles of quantum mechanics, one can devise a
physics thought experiment giving a computing machine 
which is apparently more powerful than the standard Turing machine 
model of theoretical computer science.  Using only polynomial
resources, these quantum computers can
compute certain functions which are not known
to be computable on classical digital computers
in less than exponential time \cite{Simo, Shor, Shorb}.  The potentially
most useful algorithms for quantum computers discovered
so far include prime 
factorization and simulation of certain quantum mechanical systems.  

Given these theoretical results, a natural question is whether such 
computers could ever be built.  Ingenious designs for such
computers have recently been proposed \cite{CiZo}, and currently
several experiments are underway in attempts to build small
working prototypes \cite{MoMeKiItWi}.  Even if small quantum computers can 
successfully be built, scaling these up to computers that are
large enough to yield useful computations could present
formidable difficulties.  

One of these difficulties is {\it decoherence} \cite{Land, Unru, ChLaShZu}.  
Quantum computation
involves manipulating the quantum states of objects that are
in coherent quantum superpositions.  These superpositions, however, tend
to be quite fragile and decay easily; this decay phenomenon is called
decoherence.  One way of thinking about decoherence is to consider the
environment to be ``measuring'' the state of a quantum system by
interacting with it \cite{Zure}.  

A second potential obstacle to building quantum computers is
inaccuracy \cite{Land, BeVa, BeBeBrVa}.  Quantum computers are 
fundamentally analog-type devices;
that is, the state of a quantum superposition depends on certain
continuous parameters.  For example, one of the common quantum gates
used in quantum computations is a ``rotation'' of a quantum bit by an
angle $\theta$.  When doing this transformation, there will naturally tend
to be some inaccuracy in this angle $\theta$. For the quantum computation
to successfully yield the correct result, this inaccuracy must be less 
than the amount of inaccuracy which the computation can tolerate.  
All quantum gates are potentially analog in that there will 
be some amount of inaccuracy in any physical implementation; that is,
the output quantum state will not be precisely the desired state.
What must be shown is that the tolerance of the computation to
these inaccuracies is large enough to permit quantum gates to be 
built.

These two difficulties are closely related.  Decoherence can be
expressed purely in terms of inaccuracies in the state of the quantum system
and an auxiliary quantum system interacting with it
called the ``environment.''
Thus, decoherence-reduction methods can often be used to correct 
inaccuracy, and vice versa.  
It has already been shown that
the use of quantum error correcting codes 
\cite{Shor2, CaSh, Stea, EkMa, BeDiSmWo}
can reduce both decoherence and inaccuracy dramatically
during transmission and storage of quantum data.
We build upon these techniques to show that the use of
these codes can also reduce decoherence
and inaccuracy while performing computations on quantum data.

Until now, the best estimate on the amount of inaccuracy required to
permit $t$ steps of quantum computation was $O(1/t)$ \cite{BeVa}.  
We show in this paper
that quantum circuits can be made substantially more
fault-tolerant.  For any 
polynomial size quantum circuit, we show how to construct a 
fault-tolerant version of the quantum circuit which computes the 
same function and also has polynomial size.
This circuit can tolerate 
$O(1/\log^c n)$ inaccuracy in the quantum gates, 
and decoherence averaging $O(1/\log^c n)$ per step.  

More specifically, we use the quantum
circuit model of computation, augmented with 
measurement operations during the computation.  For noise-free quantum 
circuits, these measurement operations can always be delayed until
the end of the computation \cite{AhNi}; thus previous definitions 
of quantum circuits have sometimes only permitted measurement steps 
at the end, as this model is easier to work with
and was believed to be equivalent.  It has not yet been shown
that measurement operations can be delayed until
the end of the computation for noisy quantum circuits
(although this seems plausible since noisy quantum gates might be
used to simulate measurement by noisy classical gates, which can 
in turn perform reliable classical computation \cite{vNeum,Pipp1}).
For now, it appears to be easier to provide fault-tolerance if 
measurement operations are allowed during the computation.
There is no fundamental physical reason for requiring that
measurement be delayed until the end of the computation.  

The techniques used in this paper to build fault-tolerant quantum circuits 
rely heavily on quantum error correcting codes \cite{Shor2, CaSh, Stea,
EkMa, BeDiSmWo}.   These codes can be used to encode $k$ quantum bits 
(qubits) of 
data into $n$ qubits of data so as to protect the data if errors 
occur in any $t$ of these $n$ qubits, where $n$, $k$ and $t$ are values 
which depend on the code used.  (But note that
$t$ cannot exceed some upper bound 
depending on $n$ and $k$, analogous to upper bounds in classical 
information theory \cite{CaSh, BeDiSmWo}.)  These codes were previously 
known to be potentially useful for storage and transmission
of quantum data.  It was not clear whether such codes could be used 
to prevent errors during quantum computation: more specifically,
it was not known how to compute with encoded qubits without decoding them, 
and decoding the qubits in order to compute exposes them 
to potential errors.  Further, decoding or correcting errors in quantum 
codes is in itself a quantum computation.  It was not known how to 
correct errors using noisy quantum gates without possibly 
introducing worse errors. 

Quantum error-correcting codes map qubits into blocks of qubits 
so that a small number of errors in the qubits of any block has little 
or no effect on the encoded qubits.  We find circuits for correcting 
errors in the encoding qubits and for computing with the encoded qubits 
so that if these circuits are implemented with slightly noisy gates, 
only a small number of errors result in the encoding qubits in each 
block, and thus the encoded qubits are not disturbed.

This paper shows both how to correct 
errors in encoded qubits using noisy gates and also how to compute 
on these encoded qubits without ever decoding the qubits.
We can thus alternate steps which perform computations on encoded 
qubits with steps that correct any errors that have occurred during 
the computation. If the error probability is small enough and the 
errors are corrected often enough, the probability that we have more 
errors than our quantum error correcting code can deal with 
remains small.  This ensures that we have a high probability of
completing our computation before it has been derailed by errors.
Our results thus show that if quantum gates can be made only moderately
reliable through hardware, substantial further improvements in 
reliability may be achievable through software.

At this point it may be informative to compare the classical and quantum
situations with respect to computing with noisy gates.  Classical
storage and transmission of data through noisy media can be
accomplished with relatively little overhead by using error correcting
codes.  However, performing classical computation with noisy gates 
is considerably harder.  While {\it ad hoc} methods can be used to 
reduce error in classical computers at relatively low cost, general 
techniques for performing reliable computation with noisy gates 
require a logarithmic increase in the size of classical circuits 
\cite{DoOr,Pipp2}.  These techniques involve keeping several
copies of every bit, and periodically reconciling them by 
setting them to the majority value \cite{vNeum,Pipp1,Renn}.  

Digital circuitry is reliable enough that these techniques are 
only cost-effective when reliability is of paramount importance 
\cite{Renn}.  The techniques given in this paper are even more 
costly in that they require a polylogarithmic increase in the size 
of quantum circuits.  Quantum gates, however, are inherently less
reliable than classical gates and thus in the quantum setting,
the benefit of these techniques may justify their cost.  

The techniques in this paper are also related to a different 
classical problem that occurs in cryptography.  It is possible to
perform computation on data that has been encoded and shared among several 
processors so that the data known collectively by any 
small subset of the processors gives no information about the 
unencoded data \cite{BeGoWi}.  This is similar to the quantum mechanical
requirement that measurement of the states of a small number of qubits 
in a fault-tolerant quantum circuit must give no information about
the unencoded data, and the techniques used in \cite{BeGoWi} are
similar to those used in this paper.

The rest of the paper is organized as follows.  In Section~2, we 
briefly review the quantum circuit model of quantum computation
that we use (which also allows additional measurement
steps during the computation).  In Section~3, we briefly
review the quantum error correcting codes discovered independently
by Calderbank and Shor and by Steane 
\cite{CaSh, Stea}, which are used in our construction.  We also review 
some of the theory of classical error correcting codes.
In Section~4, we show how errors
in quantum information encoded in these quantum error correcting codes 
can be corrected using
slightly noisy gates without introducing more error than
is eliminated.  In Sections~5 and 6, we show how to compute using 
encoded quantum data:  Section~5 shows how to
perform Boolean linear operations and certain
$\pi/2$ rotations and Section~6
shows how to perform Toffoli gates.  Together, these form a universal
set of gates for quantum computation.  In Section~7, we put all these
pieces together to obtain robust quantum computation and we discuss 
some open problems.

Since our submission of this paper, we have learned that Zurek and 
Laflamme have independently investigated gates that calculate on 
encoded qubits \cite{ZuLa}.  Some of their ideas might be useful 
for simplifying our constructions.

\section{Quantum circuits}

The model of quantum computation we use is the
quantum circuit model \cite{Yao}.  
Our quantum computation will be done in the quantum
state space of $n$ two-state quantum systems (e.g., spin-$\frac{1}{2}$
particles).  Each of these two-state quantum systems can be in a 
superposition of two quantum states, which we represent 
by $\ket{0}$ and $\ket{1}$.
The quantum state space of $n$ of these
particles is a $2^n$-dimensional complex
space, with $2^n$ basis vectors $\ket{b_1 b_2 \ldots b_{n}}$
indexed by binary strings of length $n$.  
A (pure) quantum state is simply a unit vector in this space.  
Each of the $n$ particles corresponds to one of the $n$ bit positions
in the indices of the $2^n$ basis vectors.  That is, a quantum state
is a sum
\begin{equation}
\sum_{b=0}^{2^{n-1}} \alpha_b \ket{b}
\end{equation}
where the $\alpha_b$ are complex numbers with $\sum_b | \alpha_b|^2 = 1$.  
Each of the $n$ particles is called a qubit.

A quantum gate on $k$ qubits is a $2^k$ by $2^k$ unitary matrix
which acts on the quantum state space of $k$ qubits.  To apply this
transformation to the quantum state space of $n$ qubits, we must
first decide which $k$ of the $n$ qubits we wish to apply it to.
We then apply the unitary transformation to these coordinate positions, 
leaving the binary values in the other coordinates untouched.
For quantum computation, we take $k$ to be some constant (such as 2 or 3).
Quantum gates on $k$ qubits then
involve the interaction of only a constant number of
quantum objects and thus are more likely to be physically realizable.
It turns out that for constant $k \geq 2$, as long as a reasonably 
powerful set of quantum gates is realizable, the functions computable 
in quantum polynomial time does not depend on $k$.  Such a set of 
quantum gates powerful enough to realize quantum computation
is called a {\it universal} set of quantum gates; It has been shown that
most sets of quantum gates are universal \cite{nine}.

In a measurement operation, we measure the value (0 or 1) of one of the
qubits.  This will project the system into a superposition of states
where this qubit has a definite value of either 0 or 1.  If we measure 
qubit $i$, the qubit will be measured as 0 with probability
\begin{equation}
\sum_{b|b_i=0} |\alpha_b|^2
\end{equation}
and 1 with probability
\begin{equation}
\sum_{b|b_i=1} |\alpha_b|^2.
\end{equation}
If $b_i$ is observed to be 0, say,
the relative values of the coefficients $\alpha_b$ are 
preserved on the states $\ket{b}$ with the $i$th bit of $b$ being 0, 
but they are renormalized so the resulting state is a unit vector.

A quantum computation is a sequence of quantum gates and measurements
on this $2^n$-dimensional quantum state
space.  In order to produce a {\it uniform} complexity class,
we need to require that this sequence can be computed by a classical
computer in polynomial time.  We allow the classical computer to
branch depending on the measurement steps; that is, after a
step which measures $b_i$, different quantum gates can be applied
depending on whether $b_i$ was observed to be 0 or 1.

For our fault-tolerant quantum circuits, we need results on
universal sets of gates for quantum computation.  One of the 
simplest universal sets contains the {\sc xor} (also called the
controlled {\sc not}) gate and all one-qubit gates \cite{nine}.  
The controlled {\sc not} gate, which maps basis states as follows:
\begin{eqnarray}
\ket{00}  &\rightarrow  &\ket{00} \nonumber \\
\ket{01}  &\rightarrow  &\ket{01} \nonumber \\
\ket{10}  &\rightarrow  &\ket{11} \nonumber \\
\ket{11}  &\rightarrow  &\ket{10} \,,
\end{eqnarray}
operates on two qubits and
negates the target qubit if and only if the control qubit is 1.
We say that this {\sc xor}s the control qubit into the target qubit.
Two canonical one-qubit gates are rotations around the $x$-axis
by an angle $\theta$, which take 
\begin{eqnarray}
\ket{0} &\rightarrow & \phantom{-}\cos(\theta/2) \ket{0} + \sin(\theta/2) 
\ket{1} \nonumber \\
\ket{1} &\rightarrow & -\sin(\theta/2) \ket{0} + \cos(\theta/2) \ket{1} 
\nonumber\,,
\end{eqnarray}
and rotations around the $z$-axis by $\phi$, which take
\begin{eqnarray}
\ket{0} &\rightarrow & \phantom{e^{i \phi /2}} \ket{0} \nonumber \\
\ket{1} & \rightarrow &  e^{i \phi /2}\ket{1} \nonumber \,.
\end{eqnarray}

We need the result that the following set of three gates ---
rotations around the $x$-axis and the $z$-axis by
$\pi/2$, and Toffoli gates --- is a
universal set of gates sufficient for quantum computation.  The proof of this
involves showing that these gates can be combined to produce a set of
gates dense in the set of 3-qubit gates.  Because of space considerations,
the details are left out of this abstract.

The Toffoli
gate is a three-qubit gate, as follows:
\begin{eqnarray}
\ket{000} &\rightarrow& \ket{000} \nonumber\\
\ket{001} &\rightarrow& \ket{001} \nonumber\\
\ket{010} &\rightarrow& \ket{010} \nonumber\\
\ket{011} &\rightarrow& \ket{011} \nonumber \\
\ket{100} &\rightarrow& \ket{100} \nonumber\\
\ket{101} &\rightarrow& \ket{101} \nonumber\\
\ket{110} &\rightarrow& \ket{111} \nonumber\\
\ket{111} &\rightarrow& \ket{110} \label{Toff} \, , 
\end{eqnarray}
The Toffoli gate is
a reversible classical gate which is universal for classical computation.
The {\sc xor} is also a classical gate, but the only classical functions that
can be constructed with it are linear Boolean functions; it takes three bits
to provide a reversible classical gate which is universal for classical
computation.  (Recall that all quantum gates must be reversible).
We will show that fault-tolerant quantum computation is possible by
showing how to do both $\pi/2$ rotations and Toffoli gates fault-tolerantly.

This paper is too short to fully discuss error models in quantum 
circuits.  We work with a simplified error model which is easy 
to analyze.  We assume that no errors occur in quantum ``wires'' 
in our circuits,
but only in the quantum gates.  In practice, unless very stable quantum
states are used to store data, quantum bits will degrade somewhat
between their output by one quantum gate and their input into another.
Practical large-scale quantum computation thus might require storage of
quantum data using error correcting codes and periodic
error correction of the memory in order to avoid excess accumulation of 
errors.  For large amounts of 
memory, this may necessitate parallel processing to keep the memory
from decaying faster than it can be accessed.

For the error model in our quantum gates, we assume for each gate
that with some probability $p$, the gate produces unreliable output, 
and with probability $1-p$, the gate works perfectly.  This model thus 
assumes ``fast'' errors, which cannot be prevented by the quantum watchdog 
(quantum Zeno) effect \cite{Peres}.  This type of error encompasses
a standard model for decoherence, where, with some small probability, 
the state of a gate is ``measured'' during its operation.
Fault-tolerant circuits which can correct ``fast'' errors are also 
able to correct ``slow'' errors.  
These include the standard model of inaccurate 
gates where the unitary matrices the quantum gates implement are not 
precisely those specified.  One way to analyze these error models 
is to use density matrices \cite{Peres, AhNi}.

\section{Error-correcting codes}
\label{sec-codes}

The construction of quantum error correcting codes relies heavily on
the properties of classical error correcting codes.  We thus 
first briefly review certain definitions and properties related to 
binary linear error correcting codes.  We only consider vectors and
codes over ${\sf F}_2$, the field of two elements, so we have
$1+1=0$.  A binary vector $v \in {\sf F}_2$ with $d$ 1's is said to 
have {\it Hamming weight} $d$.
The {\it Hamming distance} $d_H(v,w)$
between two binary vectors $v$ and $w$ is the Hamming weight of $(v+w)$.  

A {\it code} $\calC$ of length $n$
is a set of binary vectors of length $n$, called {\it codewords}.
In a {\it linear code} the codewords are those vectors in
a subspace of ${\sf F}_2^n$ (the $n$-dimensional vector 
space over the field ${\sf F}_2$ on two elements).  The {\it minimum 
distance} $d=d(\calC)$ of a binary code $\calC$ is the minimum 
distance between two distinct codewords.  If $\calC$ is linear then 
this minimum distance is just the minimum Hamming weight of a 
nonzero codeword.

A linear code $\calC$ with length $n$, 
dimension $k$ and minimum weight $d$ is called an $[n,k,d]$ code.
A generator matrix $G$ for a code $\calC$ is an $n$ by $k$ matrix
whose row space consists of the codewords of $\calC$.  
A parity check matrix $H$ for this code is an $n$ by $n-k$ matrix
such that  $H x^T = 0$ for any $x$ in the code.  In other words,
the row space of $H$ is the subspace of ${\sf F}_2$ perpendicular 
to $\calC$.   

For a code $\calC$ with minimum weight $d$,
any binary vector in ${\sf F}_2^n$ is within Hamming distance  
$t=\lfloor \frac{d-1}{2} \rfloor$ of at most one codeword; thus, a
code with minimum weight $d$ can correct $t$ errors made in the bits
of a codeword; such a code is thus said to be a $t$ error correcting 
code.  Suppose we know a vector $y$ which is a codeword with $t$ or
fewer errors.  All the information needed to correct $y$ is
contained in the {\it syndrome} vector $s = H y^T$.  If the syndrome
is 0, we know $y \in \calC$.  Otherwise, we can deduce the positions
of the errors from the syndrome.  To correct the errors, we need then
only apply a {\sc not} to the bits in error.  Computing the positions of the
errors from the syndrome is in general a hard problem; however, for 
many codes it can be done in polynomial time. 

The dual code $\calC^\perp$ of a code $\calC$ is the set of vectors
perpendicular to all codewords, that is $\calC^\perp = \{ v \in 
{\sf F}_2^n : v \cdot c = 0 \ \linebreak[3]\forall c \in C \}$.  It 
follows that if $G$ and $H$ are generator and parity check matrices
of a code $\calC$, respectively, then $H$ and $G$ are generator and
parity check matrices for $\calC^\perp$.  

Suppose we have an $[n,k,d]$ linear code $\calC$ such that 
\begin{equation}
\calC^\perp  \subset \calC \subset {\sf F}_2^n.
\label{weak-self-dual}
\end{equation}
We can use $\calC$ 
to generate a quantum error correcting code which will correct errors
in any $t = (d-1)/2$ or fewer qubits.  More details and proofs
of the properties of these codes can be found in \cite{CaSh, Stea}; 
in this abstract we only briefly describe results shown in
these papers.

We will be using two different expressions for the codewords of
our quantum codes.  The first is described in \cite{Stea}.  Suppose
that $v \in \calC$.  We obtain a quantum state on $n$ qubits
as follows:
\begin{equation}
\ket{s_v} = 2^{-(n-k)/2} \sum_{w \in \calC ^\perp} \ket{v+w}\,.
\end{equation}
We refer to this as the $s$-basis, and will be using it in most
of our calculations.
It can easily be shown that $\ket{s_v} = \ket{s_u}$ if and only if
$u + v \in \calC^\perp$.  Thus, there are $\dim{\calC} - \dim{\calC^\perp}
= 2k-n$ codewords in our quantum code.

By rotating each of the $n$ qubits in a quantum codeword as follows:
\begin{eqnarray}
\ket{0} & \rightarrow & {\textstyle\frac{1}{\sqrt{2}}}(\ket{0}+\ket{1} )
\label{eq-rotation} \nonumber \\
\ket{1} & \rightarrow & {\textstyle\frac{1}{\sqrt{2}}}(\ket{0}-\ket{1} )
\,,
\end{eqnarray}
it is easy to verify that we obtain the quantum state
\begin{equation}
\ket{c_v} = 2^{-k/2} \sum_{w \in \calC} (-1)^{v\cdot w} \ket{w}\,.
\end{equation}
As with the $s$-basis, it is easily checked that $\ket{c_v} = \ket{c_u}$
if and only if $u+v \in \calC^\perp$.
We call this the $c$-basis for our code; 
it was first described in \cite{CaSh}.

In quantum codes, in order to correct $t$ arbitrary errors, all that is 
needed is be able to correct any $t$ errors of the following three types
\cite{BeDiSmWo, EkMa}:
\begin{itemize}
\item[1)] bit errors, where $\ket{0} \rightarrow \ket{1}$ and $\ket{1}
\rightarrow \ket{0}$ for some qubit,
\item[2)] phase errors, where $\ket{0} \rightarrow \ket{0}$ and $\ket{1}
\rightarrow -\ket{1}$ for some qubit,
\item[3)] bit-phase errors, where $\ket{0} \rightarrow \ket{1}$ and $\ket{1}
\rightarrow -\ket{0}$ for some qubit.
\end{itemize}
Phase errors can be converted to bit errors and vice versa by applying
the change of basis in Equation (\ref{eq-rotation}).  Note that in both 
the $s$- and the $c$-bases, the codewords are superpositions of codewords 
in our error correcting code $\calC$.   Thus, to correct errors
in the above quantum error correcting code, we can first correct bit
errors in the $s$-basis using classical error correcting techniques, 
change bases as in (\ref{eq-rotation}), and then correct phase errors in 
the $s$-basis (which have become bit errors in the $c$-basis)
by using the same classical error correcting code techniques.  It remains
only to show that these two correction steps do not interfere with each
other; this is done in detail in \cite{CaSh}.

\section{Correcting errors fault-tolerantly}
\label{sec-correct}

To correct errors in quantum codes, we first correct errors in 
the codewords seen in the $\ket{s}$
basis; we next rotate each qubit of the code as in Equation 
(\ref{eq-rotation})  (this could be done symbolically);
and finally we correct the errors in the $\ket{c}$ basis.  In either
basis, the error 
correction can be done by determining which bits are in
error by first computing the (classical) syndrome and then
applying a {\sc not} operation (in the appropriate
basis) to those bits which are in error.  The hard part 
of this procedure turns out to be the
determination of which bits are in error.

Recall that to determine which bits are in error in classical codes, 
we need to compute the syndrome $S = H x^T$, where $x$ is the word
we are trying to decode.  This procedure
will work equally well for our quantum codes, although in this case
it must be done separately for the bit and the phase errors.
It is this step which is hard to do fault-tolerantly.
Computing the error from the
syndrome is hard for arbitrary linear codes, but if we have
made a good choice of our classical code $\calC$, then we have
a polynomial-time algorithm for computing the error from the syndrome.
Note that since we can measure the syndrome, this decoding step
can be done on a classical computer.
We discuss later in Section~7
classical codes which are both polynomial-time
decodable and also strong enough to use to construct 
fault-tolerant quantum circuits.  

The obvious way to measure the syndrome would be the following.  For
each row of the parity check matrix $H$, we take an ancillary 
qubit (or {\it ancilla}) which starts in the state $\ket{0}$, perform 
a controlled {\sc not} from each of the qubits corresponding to
1's in that row of the parity check matrix into the ancillary
qubit, and finally measure this ancillary qubit.  Assuming that 
we do not make any errors in our quantum calculations, this works perfectly.
Unfortunately, this method is not robust against quantum errors.  
One possible quantum error is a measurement error, where the
qubit is measured as $\ket{0}$ instead of $\ket{1}$ or vice versa.  
If this were the only kind of error possible, 
it could be controlled by repeating 
the measurement several times.  Unfortunately, this method also permits
much worse types of error.

Suppose that we apply the above method, and halfway through the process
(i.e., when we have {\sc xor}ed half of the 1's in that row of the parity
check matrix $H$ into our ancillary qubit), the state of the ancilla changes
spontaneously.  We now quite possibly obtain a wrong value for that 
bit of the syndrome, but much worse, we have also changed the state 
of the quantum codeword.  This can most easily be seen by changing the 
basis for all the qubits in the codeword, as well as for our ancilla, by
Equation (\ref{eq-rotation}).  The effect of this change of basis
on an controlled {\sc not} is to reverse
the roles of the control and the target qubits.  Thus, in the rotated basis,
our computation {\sc xor}s the ancillary
qubit \mbox{$(\ket{0} + \ket{1})/\sqrt{2}$} into certain qubits of our 
codeword.
If everything proceeds without error, this measurement changes the state
of our quantum codeword by {\sc xor}ing $\ket{000\ldots 00} + \ket{111\ldots 11}$ 
into the qubits corresponding to 1's in some
row of the parity check matrix $H$.  Since $H$ is a generator matrix for 
the dual code $\calC^\perp$,
it follows after some calculation
that this does not change our codeword.  However, if the state of
the ancillary qubit changes in the middle of our computation,
we could end up {\sc xor}ing $\ket{000\ldots 11} + \ket{111\ldots 00}$ into
our quantum codeword.  In this scenario, one error during the quantum
error correction would possibly lead to more errors in our quantum
codeword than it is possible to correct.  This technique thus cannot
be used for error correction in quantum computation.

We use a slightly different technique to measure the syndrome
without introducing too many errors in our quantum codewords.
To measure the $i$th bit of the syndrome, we first construct a 
``cat state''
\begin{equation}
{\textstyle\frac{1}{\sqrt{2}}} (\ket{000\ldots 0}+ \ket{111\ldots 1})
\end{equation}
where the number of qubits in this state (say $l$)
is equal to the number of 1's
in the $i$th row of our parity check matrix $H$.
(This is called a ``cat state'' after Schr\"odinger's renowned
cat, as it is the one of the most unstable states of $l$ qubits.)
We next {\it verify} the cat state by measuring the {\sc xor} of
random pairs of its qubits (this can be done using an auxiliary 
qubit). 
If all these
measurements are 0, this will ensure that the cat state is a
superpositions of states containing nearly all 0's or nearly all 1's,
although the relative phase of the all-0's and the all-1's
states may still be in error.  
If these measurements are not all 0's, we construct another cat state
and repeat the process.

To use the cat state, we next
rotate each qubit of the cat state as in transformation
(\ref{eq-rotation}).  If we do not make any errors, this gives a
state
\begin{equation} 
2^{-(l-1)/2} \sum_{b:b\cdot\bar{1} = 0} \ket{b}
\end{equation} 
where $l$ is the number of qubits in our cat state and $\bar{1}$ is
the length $l$ all-ones vector.  In other words, this is the superposition of
all states with an even number of 1's.  Finally, we {\sc xor} each of the qubits
of the $i$th row of the parity check matrix into one of the qubits of the
rotated cat state.  Since this rotated cat state was in the superposition
of all even-parity states, if we now measure the qubits in this state,
the parity of number of 1's observed will be the $i$th bit of the syndrome.
More important, even if we have made $r$ errors
in our calculation, the back action of the {\sc xor}s on the encoded
state will not introduce more than $r$ errors in qubits of our codeword.  
Thus, we can measure bits of the syndrome and keep our encoded states 
well protected.  This allows us to correct errors while introducing 
on the average fewer errors than we correct.

Measurement of the syndrome using the method above is not guaranteed
to give the right answer.  What it does guarantee is that 
errors in the measurement operation are unlikely to produce
catastrophic back action which destroys the encoded state 
beyond repair.  We still need to ensure somehow
that we obtain the right value for the error syndrome before attempting 
to correct the errors.  One way to get the right error syndrome with
probability $1-1/t$ is to repeat the above measurement $O(\log t)$ times.
If we obtain the same error syndrome each time, the probability that we
have made the same error repeatedly is very small.  If we obtain
different syndromes, we can keep repeating the measurement until
the same error syndrome is obtained $O(\log t)$ times in a row.  If the
error rate is set low enough this guarantees that we correct 
the error with probability at least $1 - 1/t$,

\section{XOR Gates and {\boldmath $\frac{\pi}{2}$} Rotations}
\label{sec-linear}

In order to give our construction for fault-tolerant quantum circuits
in detail, we first need to introduce more facts about error correcting
codes.  These can be found explained in more detail in coding theory 
books (such as \cite{MaSl}).  We will be using
codes with $\dim(\calC) - \dim(\calC^\perp) = 1$; they are thus rather
inefficient in that they code one qubit into $n$.  The codes we use
can be constructed from self-dual codes with $\calC = \calC^\perp$
by deleting any one coordinate; such codes are called {\it punctured}
self-dual codes.  If a code $\calC$ has minimum distance $d$, the 
punctured code has minimum distance at least $d-1$ and can thus
correct $\lfloor d/2 -1 \rfloor$ errors.  Binary self-dual codes have
the property that all codewords have an even number of ones, since
every codeword must be perpendicular to itself.  Some binary self-dual codes 
have the additional property that the number of ones in all codewords is
divisible by 4.  These are called {\it doubly even} codes and their
properties are useful in constructing fault-tolerant quantum 
circuits.

Suppose that we have a punctured self-dual code $\calC$ with length
$n$, dimension $k=(n+1)/2$ and minimum distance $d$.  Consider the
corresponding quantum codewords $\ket{s_v}$.  From the previous section,
we have that the number of different
quantum codewords is $2k-n = 1$.  It is easy to verify that one of these
consists of the superposition of all codewords in $\calC$ with
even weight (these are the codewords of $\calC^\perp$) and the other 
consists of the superposition of all codewords of $\calC$ of odd
weight.  We label these $\ket{s_0}$ and $\ket{s_1}$ respectively.
It is also easy to see that the $c$-basis of our code looks like:
\begin{eqnarray}
\ket{c_0} &=& {\textstyle\frac{1}{\sqrt{2}}} (\ket{s_0} + \ket{s_1}),
\nonumber\\
\ket{c_1} &=& {\textstyle\frac{1}{\sqrt{2}}} (\ket{s_0} - \ket{s_1}) 
\,.
\end{eqnarray}

Recall that we get from the $s$-basis to the $c$-basis of our quantum
codes by applying the transformation (\ref{eq-rotation}) to each qubit
of our codewords. It is clear from the above equations
that applying this transformation to each qubit individually
also applies this
transformation to the encoded states.  Further, this transformation
is fault-tolerant.  Suppose that there were at most $r$ errors in qubits
of the input state.  This transformation is applied separately to
each qubit, so the output state will also contain $r$ errors in it
if the transformation was applied perfectly.  Even if the 
transformation is imperfect, 
an error in one application of a quantum gate can only
affect one qubit since the transformation is applied bitwise, so
even with noisy gates, this transformation can
introduce only a small number of errors, 

A number of other transformations of the encoded qubits 
can also be performed by applying them bitwise to the codeword.
We demonstrate this
with an {\sc xor} gate.  Suppose we have two different qubits encoded, 
$\ket{s_a}$ and $\ket{s_b}$.  Expanding these, we have the quantum state
\begin{equation}
\ket{s_a}\ket{s_b} = 
{\textstyle 2^{-(n-k)}}
\sum_{w \in \calC^\perp} \ket{w+a} 
\sum_{w' \in \calC^\perp} \ket{w'+b} 
\end{equation}
Applying an {\sc xor} from the $i$th qubit of the first codeword into the $i$th qubit
of the second codeword, we obtain
\begin{eqnarray}
{\textstyle 2^{-(n-k)}}
\sum_{w \in \calC^\perp} \Big(\ket{w+a} 
\sum_{w' \in \calC^\perp} \ket{w'+b+w+a} \Big). 
\end{eqnarray}
If $w'$ ranges over all codewords in $\calC^\perp$, then for
any fixed $w \in \calC^\perp$, $w'+w$ also ranges over all codewords in
$\calC^\perp$.  Thus, the above sum can be rewritten as 
$\ket{s_a}\ket{s_{a+b}}$.  This operation works for any $\calC$ with
$\calC^\perp \subset \calC$ as in
Equation~(\ref{weak-self-dual}).  For punctured self-dual codes $\calC$,
with $\dim{\calC^\perp} =
\dim{\calC} - 1$, this gives an {\sc xor} gate.

Other operations which can also be done bitwise are the phase change
operation
$\ket{s_a}\ket{s_b} \rightarrow (-1)^{a\cdot b} \ket{s_a}\ket{s_b}$ and
(for punctured doubly even self-dual codes), the rotation 
\begin{eqnarray}
\ket{s_0} & \rightarrow  & \phantom{i} \ket{s_0} \nonumber \\
\ket{s_1} & \rightarrow  & i \ket{s_1} \,.
\end{eqnarray}
The calculations for both of these cases are straightforward.

These above operations which can be done bitwise are not enough to
provide a universal set of gates for quantum computation.  They
only generate unitary matrices in the {\it Clifford
group}, which is a finite
group of unitary transformations in $2^n$-dimensional
complex space that arises in several areas of mathematics
\cite{Kerdock}.   The 
transformations in this group corresponding to classical
computation are the {\it linear Boolean} 
transformations, which can be built out of {\sc xor} and {\sc not} gates.  To
obtain a set of gates universal for quantum computation, we
add the Toffoli gates as in 
Equation (\ref{Toff}).  This construction is discussed in the next section. 

\section{Toffoli gates}
\label{sec-Toff}

We construct our Toffoli gate in two stages.  We first
show how to construct a fault-tolerant
Toffoli gate given a set of ancillary quantum 
bits known to be in the encoded state
$\frac{1}{2}(\ket{s_0 s_0 s_0} + \ket{s_0 s_1 s_0} + \ket{s_1 s_0 s_0} + 
\ket{s_1 s_1 s_1})$.  This procedure is done using only linear Boolean
operations and $\pi/2$ rotations on the encoded qubits.  We next
show how to fault-tolerantly put a set of ancillary qubits
into the above state.  This operation will be somewhat harder, as
it cannot be done using operations in the Clifford group.  To construct 
this state, we use a ``cat'' state $(\ket{000 \ldots 0} +
\ket{111 \ldots 1})/\sqrt{2}$ as we did in Section~4.

A technique used both in this section and in Section~4
is that of first constructing an ancillary set of qubits 
known to be in a certain state and then using them to perform 
operations on another set of qubits.  This is 
reminiscent of techniques used in several quantum communication 
papers.  In quantum teleportation \cite{BeBrCrJoPeWo}, if two 
researchers
share an EPR pair, they can use this pair and classical communication
to ``teleport'' the quantum state of a particle from one researcher 
to another.
In \cite{BeBrPoScSmWo}, a small number of ``USDA'' pairs of 
qubits known to be in perfect EPR states can used to purify a 
set of noisy EPR states, sacrificing some of them
but yielding a large set of good EPR pairs.  This 
paradigm may prove useful in other quantum computations.

\subsection{ Using the ancillary state}

Suppose we had an ancillary set of qubits known to be in the encoded state 
\begin{equation}
\ket{A} = \frac{1}{2}(\ket{s_0 s_0 s_0} + \ket{s_0 s_1 s_0} + \ket{s_1 s_0 s_0} + 
\ket{s_1 s_1 s_1}).  
\label{Toff-ancilla}
\end{equation}
We now show how to use these to make a
Toffoli gate on 3 other encoded qubits, using Boolean linear operations
and $\pi/2$ rotations.  
Recall that the Toffoli gate transforms qubits by
negating the third qubit if and only if the first two are 1's.  
We first build a gate that makes the following transformation 
taking two encoded qubits to three encoded qubits.  
\begin{eqnarray}
\ket{s_0s_0} &\rightarrow& \ket{s_0s_0s_0} \nonumber\\
\ket{s_0s_1} &\rightarrow& \ket{s_0s_1s_0} \nonumber \\
\ket{s_1s_0} &\rightarrow& \ket{s_1s_0s_0} \nonumber\\
\ket{s_1s_1} &\rightarrow& \ket{s_1s_1s_1} \label{twobitToff2}\,.
\end{eqnarray}
Note that this gate adds a third (encoded) qubit, which is a 1 if and only
if the first two are both 1's, and which is
0 otherwise.  This gate uses the ancillary state 
$\ket{A}$ described above, as well as linear operations,
which can be performed robustly as in Section~5.

To perform these transformations,
we first append the ancilla $\ket{A}$ to the
first two qubits.
We next {\sc xor} the third qubit into the first, and the fourth qubit into
the second.  This produces the transformation
\begin{eqnarray}
\ket{s_0s_0}\ket{A} &\rightarrow&{\textstyle \frac{1}{2}} \big(
\ \ket{s_0 s_0 s_0 s_0 s_0} + \ket{s_0 s_1 s_0 s_1 s_0} \nonumber \\
& & \quad + 
\ket{s_1 s_0 s_1 s_0 s_0} + \ket{s_1s_1s_1s_1s_1}\ \big) \nonumber\\
\ket{s_0s_1}\ket{A} &\rightarrow&{\textstyle \frac{1}{2}} \big(
\ \ket{s_0 s_1 s_0 s_0 s_0} + \ket{s_0 s_0 s_0 s_1 s_0}  \nonumber \\
& & \quad +
\ket{s_1 s_1 s_1 s_0 s_0} + \ket{s_1 s_0s_1s_1s_1}\ \big) \nonumber \\
\ket{s_1s_0}\ket{A} &\rightarrow&{\textstyle \frac{1}{2}} \big(
\ \ket{s_1 s_0 s_0 s_0 s_0} + \ket{s_1 s_1 s_0 s_1 s_0} \nonumber \\
& & \quad + 
\ket{s_0s_0s_1 s_0 s_0} + \ket{s_0s_1s_1s_1s_1}\ \big) \nonumber\\
\ket{s_1s_1}\ket{A} &\rightarrow&{\textstyle \frac{1}{2}}\big( 
\ \ket{s_1 s_1 s_0s_0s_0} + \ket{s_1 s_0 s_0 s_1 s_0} \nonumber \\
& & \quad + 
\ket{s_0 s_1 s_1 s_0 s_0} + \ket{s_0 s_0 s_1s_1s_1}\ \big) \nonumber \,. \\
& & 
\end{eqnarray}
Finally, we measure the first and second encoded qubits.

Suppose 
we measure them to be $\ket{s_0s_0}$, so both encoded qubits are 0.  
Focusing on the elements of the superposition where the 
first and second encoded qubits are both 0, we get the transformation 
given by (\ref{twobitToff2}),
which is what we wanted in the first place.  Suppose, however, that
we measure the first and second 
encoded qubits to be in the state $\ket{s_0s_1}$.
Pulling out the relevant elements of the superposition, we get
\begin{eqnarray}
\ket{s_0s_0} &\rightarrow& \ket{s_0s_1s_0} \nonumber\\
\ket{s_0s_1} &\rightarrow& \ket{s_0s_0s_0} \nonumber\\
\ket{s_1s_0} &\rightarrow& \ket{s_1s_1s_1} \nonumber\\
\ket{s_1s_1} &\rightarrow& \ket{s_1s_0s_0} \,.
\end{eqnarray}
This transformation can be converted to the one we want by first applying
a controlled {\sc not} from the first qubit to the third qubit, and then applying
a {\sc not} to the second qubit.  These are both Boolean linear operations and
so can be applied fault-tolerantly.  
Putting these transformations together, we get
\begin{eqnarray}
\ket{s_0s_0} \rightarrow \ket{s_0s_1s_0}   \rightarrow  
\ket{s_0s_1s_0} \rightarrow  \ket{s_0s_0s_0} \phantom{.} &&\nonumber\\
\ket{s_0s_1} \rightarrow \ket{s_0s_0s_0}  \rightarrow 
\ket{s_0s_0s_0} \rightarrow  \ket{s_0s_1s_0} \phantom{.} &&\nonumber\\
\ket{s_1s_0} \rightarrow \ket{s_1s_1s_1}  \rightarrow 
\ket{s_1s_1s_0} \rightarrow  \ket{s_1s_0s_0} \phantom{.} &&\nonumber\\
\ket{s_1s_1} \rightarrow \ket{s_1s_0s_0}  \rightarrow 
\ket{s_1s_0s_1} \rightarrow  \ket{s_1s_1s_1}. &&
\end{eqnarray}
It is easy to check that the other two cases (where we observe
$\ket{s_1s_0}$ or $\ket{s_1s_1}$), can also be corrected to the 
desired gate by linear operations.  Thus, by observing 
two of these five qubits and applying linear operations, we have 
achieved what is nearly a Toffoli gate.

We still need to show how to get the complete Toffoli gate on
three qubits, as in Equation (\ref{Toff}).  To do this, we start
with three qubits to which we want to apply the Toffoli gate,
and apply transformation (\ref{twobitToff2}) to the
first two.  We next
apply a controlled {\sc not} from our original third qubit (represented in
fourth place below) to the newly introduced qubit (represented in
third place below).   We 
finally apply 
$\ket{s_0} \rightarrow (\ket{s_0}+\ket{s_1})/\sqrt{2}$,
$\ket{s_1} \rightarrow (\ket{s_0}-\ket{s_1})/\sqrt{2}$ to the original
third qubit (represented in fourth place below).  This
gives the transformation:
\begin{eqnarray}
\ket{s_0s_0s_0} &\rightarrow& \textstyle \frac{1}{\sqrt{2}}
\ket{s_0s_0s_0}\big(\ket{s_0}+\ket{s_1}\big)
 \nonumber\\
\ket{s_0s_1s_0} &\rightarrow& \textstyle \frac{1}{\sqrt{2}}
\ket{s_0s_1s_0}\big(\ket{s_0}+\ket{s_1}\big)
 \nonumber\\
\ket{s_1s_0s_0} &\rightarrow& \textstyle \frac{1}{\sqrt{2}}
\ket{s_1s_0s_0}\big(\ket{s_0}+\ket{s_1}\big)
 \nonumber\\
\ket{s_1s_1s_0} &\rightarrow& \textstyle \frac{1}{\sqrt{2}}
\ket{s_1s_1s_1}\big(\ket{s_0}+\ket{s_1}\big)
 \nonumber\\
\ket{s_0s_0s_1} &\rightarrow& \textstyle \frac{1}{\sqrt{2}}
\ket{s_0s_0s_1}\big(\ket{s_0}-\ket{s_1}\big)
 \nonumber\\
\ket{s_0s_1s_1} &\rightarrow& \textstyle \frac{1}{\sqrt{2}}
\ket{s_0s_1s_1}\big(\ket{s_0}-\ket{s_1}\big)
 \nonumber\\
\ket{s_1s_0s_1} &\rightarrow& \textstyle \frac{1}{\sqrt{2}}
\ket{s_1s_0s_1}\big(\ket{s_0}-\ket{s_1}\big)
 \nonumber\\
\ket{s_1s_1s_1} &\rightarrow& \textstyle \frac{1}{\sqrt{2}}
\ket{s_1s_1s_0}\big(\ket{s_0}-\ket{s_1}\big)
\,. 
\end{eqnarray}
We now observe the fourth qubit in the expression above.  If we observe
$\ket{s_0}$, it is easy to see that we have performed a Toffoli gate.
If we observe $\ket{s_1}$, we need to fix the resulting state up.  This
can be done by applying the transformation 
\begin{eqnarray}
\ket{s_as_bs_c} \rightarrow (-1)^{a\cdot b} (-1)^c \ket{s_as_bs_c}
\end{eqnarray}
to the three remaining encoded qubits (recall that
we measured the last qubit).
This is the composition of the two linear operations
$\ket{s_as_b} \rightarrow (-1)^{a\cdot b}$ and 
$\ket{s_c} \rightarrow (-1)^c$,
so it can be done fault-tolerantly using the methods of 
Section~5.

\subsection{Constructing the ancillary state}
For the last piece of our algorithm, we need to show how to construct
the ancillary state $\frac{1}{2}(\ket{s_0s_0s_0} +\ket{s_0s_1s_0} 
+\ket{s_1s_0s_0} +\ket{s_1s_1s_1})$ fault-tolerantly.   To obtain this, 
we use the technique we used in Section~4 
of introducing a ``cat'' state 
$(\ket{000 \ldots 0} + \ket{111 \ldots 1})/\sqrt{2}$, 
which we can check to ensure that it is in a state close to the 
desired state before we use it.  We use two states on $3n$ 
qubits, the state $\ket{A}$ defined in (\ref{Toff-ancilla})
and the state $\ket{B}$, which are as follows:
\begin{eqnarray}
\ket{A} &=& {\textstyle \frac{1}{2}}
( \ket{s_0s_0s_0} +\ket{s_0s_1s_0} +\ket{s_1s_0s_0} +\ket{s_1s_1s_1})
\nonumber \\
\ket{B} &=& {\textstyle \frac{1}{2}}
( \ket{s_0s_0s_1} +\ket{s_0s_1s_1} +\ket{s_1s_0s_1} +\ket{s_1s_1s_0})
\nonumber \,.
\end{eqnarray}
Note that $\ket{A}$ is the state we want, and $\ket{B}$ is easily convertible
to it 
by applying a {\sc not} operation to the third encoded qubit.  Note also that
\begin{equation}
{\textstyle\frac{1}{\sqrt{2}}}( \ket{A} + \ket{B}) =
{\textstyle\frac{1}{2\sqrt{2}}}
(\ket{s_0}+\ket{s_1}) (\ket{s_0}+\ket{s_1}) (\ket{s_0}+\ket{s_1}) \,,
\end{equation}
which is easily 
constructible through operations discussed
in Section~5. 
Let us now 
introduce some more notation: let $\ket{\bar{0}} = \ket{000\ldots0}$ and
$\ket{\bar{1}} = \ket{111\ldots1}$, where these states are both
on $n$ qubits.
Thus, $(\ket{\bar{0}} + \ket{\bar{1}})/\sqrt{2}$ is the cat state described above.
Suppose we could apply a transformation that changed states as follows:
\begin{eqnarray}
\ket{\bar{0} }\ket{A} & \rightarrow & \phantom{-}  
\ket{\bar{0} }\ket{A} \nonumber \\
\ket{\bar{1} }\ket{A} & \rightarrow & \phantom{-}  \ket{\bar{1} }\ket{A} 
\nonumber \\
\ket{\bar{0} }\ket{B} & \rightarrow & \phantom{-}  
\ket{\bar{0} }\ket{B} \nonumber \\
\ket{\bar{1} }\ket{B} & \rightarrow &  - \ket{\bar{1} }\ket{B} 
\label{ABtransform}\,,
\end{eqnarray}
We would then be able to construct state $\ket{A}$ as follows.  First,
construct
\begin{equation}
{\textstyle\frac{1}{2}}
(\ket{\bar{0}} +  \ket{\bar{1}}) (\ket{A} + \ket{B}) \,,
\label{ABstate}
\end{equation}
where the cat state has been verified as in Section~4
to make sure it is close to the
desired cat state (i.e., nearly all the probability amplitude is 
concentrated in states with either nearly all 0's or nearly all 1's).
We next apply the transformation (\ref{ABtransform}) to obtain the state
\begin{equation}
{\textstyle\frac{1}{2}}(\ket{\bar{0}} + \ket{\bar{1}}) \ket{A} +
{\textstyle\frac{1}{2}}(\ket{\bar{0}} - \ket{\bar{1}}) \ket{B} \,.
\end{equation}
Finally, we observe whether the ``cat qubits'' are in the state
$\ket{\bar{0}} \pm \ket{\bar{1}}$.  This tells us whether the unobserved
qubits contain $\ket{A}$ or $\ket{B}$.  

The probability of being in state
$\ket{A}$ can be estimated using the probabilities of error during
the quantum calculation.  If this probability is not high enough, 
we repeat this step, applying the transformation (\ref{ABtransform})
not to the state (\ref{ABstate}) but to the output state from the 
previous step along with a newly constructed ``cat state.''
Repeating this step logarithmically many
times can be shown to increase this probability of being in state
$\ket{A}$ to polynomially close to 1.

To apply the transformation (\ref{ABtransform}) to a superposition
$\ket{\bar{a}} \ket{s_b} \ket{s_c} \ket{s_d}$, it is sufficient to
apply bitwise the operation
\begin{equation}
\ket{{a_i}} \ket{b_i} \ket{c_i} \ket{d_i} \rightarrow
(-1)^{a_i(b_ic_i+d_i)} \ket{a_i} \ket{b_i} \ket{c_i} \ket{d_i} 
\end{equation}
to the $i$th qubit of 
$\ket{\bar{a}}$, $ \ket{s_b}$, $ \ket{s_c}$  and $ \ket{s_d}$ for $1 \leq i 
\leq n$.  This operation is easily accomplished by elementary quantum
gates and as it is a bitwise operation, it is fault-tolerant.

The one piece of the computation which we have not yet described is
how to construct an encoded state $\ket{s_0}$.  This can be done by
techniques similar to those described in this paper, but there is
no space to describe this in detail.

\section{Conclusions}
\label{sec-conclusions}

We now estimate the accuracy required to make
quantum circuits fault-tolerant with these methods.  The only large
binary self-dual codes I know of which are also 
decodable in polynomial time
are the $[2^{m+1},2^m,2^{m/2}]$ self-dual Reed-Muller codes
\cite{MaSl}.  These codes are indeed doubly even, but
unfortunately their minimum distance (and 
error correction capacity) grows as the square root of their length~$n$.
This will be enough for our purposes, but these codes are substantially
worse than both the theoretical maximum and than the
known constructions without all the required properties.

In order to do $s$ steps of quantum computation with a low probability
of failure, we need a quantum code which can correct $O(\log s)$
errors.  Using Reed-Muller codes, this means we need codewords
of length $O(\log^2 s)$.  In measuring the syndrome, to be
relatively sure that we have not made an error in computing it,
we measure each bit of the syndrome $O(\log s)$ times (this is
probably overkill).  Using the
measuring technique described in Section~4, even if
we make errors while measuring the syndrome, we do not substantially
affect our encoded qubits.  

Computing the error syndrome requires
a number of quantum gates proportional to the number of 1's in the
parity check matrix, which in this case is $O(\log^3 s)$.  Since we 
measure the syndrome $O(\log s)$ times in our correction step, the 
entire correction step takes $O(\log^4 s)$ operations.  We need to set
the error rate low enough so that there will be less than one error on
the average throughout this process, which means we must have
error rate less than $O(1/\log^4 s)$ per gate operation.  
Computation operations on encoded qubits take at most
$O(\log^4 s)$ steps, so these can also be 
accomplished while ensuring that there is a very small probability 
of making more errors than we can correct.

One additional result which would be nice is the discovery of better 
binary self-dual error correcting codes which are 
also efficiently decodable; 
this could substantially increase the asymptotic efficiency of the 
fault-tolerant quantum circuits described in this paper.  Another 
interesting result would be a method for performing rotations on encoded 
bits directly, rather than going through Toffoli gates.  This could be 
accomplished using techniques similar to those in Section~6
if the ancillary state $\cos(\theta) \ket{s_0} + \sin(\theta) \ket{s_1}$ could
be constructed fault-tolerantly for arbitrary $\theta$.  

The techniques in
this paper pay a moderate penalty in both space and time for making
quantum circuits fault-tolerant.  An interesting question is
how much time
and space are really required.  The space, for example, could likely be
reduced significantly by using more efficient quantum error correcting 
codes for memory.  
Another interesting question is how much noise in quantum gates can 
be tolerated while still permitting
quantum computation.  A lower bound on this quantity is shown in
\cite{AhBe}.  Better 
error-correcting codes could possibly increase the maximum allowed
error rate considerably from $O(1/\log^4 t)$, but it appears that to
get results better than $O(1/\log t)$, substantially different
techniques may be required.

Finally, the analysis in this paper is purely asymptotic.  
An analysis needs to be done to see how much fault-tolerance these 
techniques provide for quantum computations using specific numbers
of gates, and at what cost in space and time this could be accomplished.

\section*{Acknowledgements}
{
I would like to thank Rolf Landauer and Bill Unruh for their skepticism,
which in part motivated this work.  I would also like to thank David DiVincenzo
for insightful discussions on quantum fault tolerant computing, David
DiVincenzo and David Johnson for helpful
comments on earlier versions of this paper, and 
Rob Calderbank and Neil Sloane for helpful discussions on
classical error correcting codes.
}

{

}

\end{document}